# Rethinking Review Citations: Impact on Scientific Integrity

Jesús S. Aguilar-Ruiz[1]*

[1]School of Engineering, Pablo de Olavide University, 41013-Seville, Spain.
*Corresponding author. Email: aguilar@upo.es

**Abstract:** The proliferation of surveys and review articles in academic journals has impacted citation metrics like impact factor and h-index, skewing evaluations of journal and researcher quality. This work investigates the implications of this trend, focusing on the fields of Computer Science, where a notable increase in review publications has led to inflated citation counts and rankings. While reviews serve as valuable literature overviews, they should not overshadow the primary goal of research —to advance scientific knowledge through original contributions. We advocate for prioritizing citations of primary research in journal articles to uphold citation integrity and ensure fair recognition of substantive contributions. This approach preserves the reliability of citation-based metrics and supports genuine scientific advancement.

The central aim of research is to expand the boundaries of human knowledge by uncovering new facts, theories, and insights. Historically, the term research (derived from Old French "recerche") has evolved from meaning "careful or diligent search" to denote a structured and methodical process of inquiry. Today, research encompasses both the pursuit of new discoveries and the systematic refinement of existing knowledge through literature reviews, hypothesis testing, and experimentation. This ongoing cycle aims to deepen our understanding of the world, contributing substantially to societal advancement [Ioannidis, 2014].

Research, when successful, is typically published as an article. Journals are the primary medium for such publications. Following the work of Eugene Garfield [Garfield, 1955], journals began to be ranked by their impact factor. Ideally, higher-quality articles should be published in higher-ranking journals. Reviewers assess whether an article meets the quality standards of a journal, while authors indirectly influence a journal's ranking by including citations to its papers.

Nonetheless, this process is subject to significant biases. Authors frequently employ rhetorical citations [Bao et Teplitskiy, 2023], such as reviews, surveys, or even unrelated works, and reviewers might suggest citing specific articles, sometimes favoring their own work or preferred journals. Editors might encourage citations that enhance their journal's metrics. As a result, citation metrics, which are intended to reflect research impact, can become skewed by these practices —authors aim to enhance their h-index, while journals strive to boost their impact factor.

Prominent research naturally attracts more attention from other scholars, leading to higher citation counts and better journal rankings. Nevertheless, over the last two decades this process has become distorted due to the undue influence of surveys and reviews on impact factor calculations.

While surveys or reviews can serve as a foundation in the research process, they should not be the goal of research, as their direct contribution to scientific progress is minimal. Typically, Ph.D. theses include a chapter dedicated to literature reviews and state-of-the-art analysis, but they do not end there, as these sections do not introduce new ideas or scientific advancements. Although such reviews are valuable for providing an overview and context, they should not be considered as significant original contributions. They are often already included in the thesis document and made accessible through institutional repositories.

The quality of a journal, measured by the overall quality of the papers it publishes, should not be influenced by the inclusion of surveys or reviews. Over the years, research quality has been assessed by state research agencies based on factors like the quantity and quality of publications. As a result, researchers aim to publish in high-ranking journals and to accumulate citations. Recently, the proliferation of review articles has somewhat compromised the perceived quality of journals and the standing of researchers. Journals with more review articles tend to achieve higher rankings, while researchers who publish these articles often gain inflated indicators of research impact, such as a higher h-index.

For example, in Computer Science, the number of journal articles published annually remained below one hundred thousand until 2001, with surveys and reviews accounting for less than 1%. After 2001, the total number of articles started to grow, surpassing six hundred thousand in 2022, with surveys and reviews exceeding 2.5%.

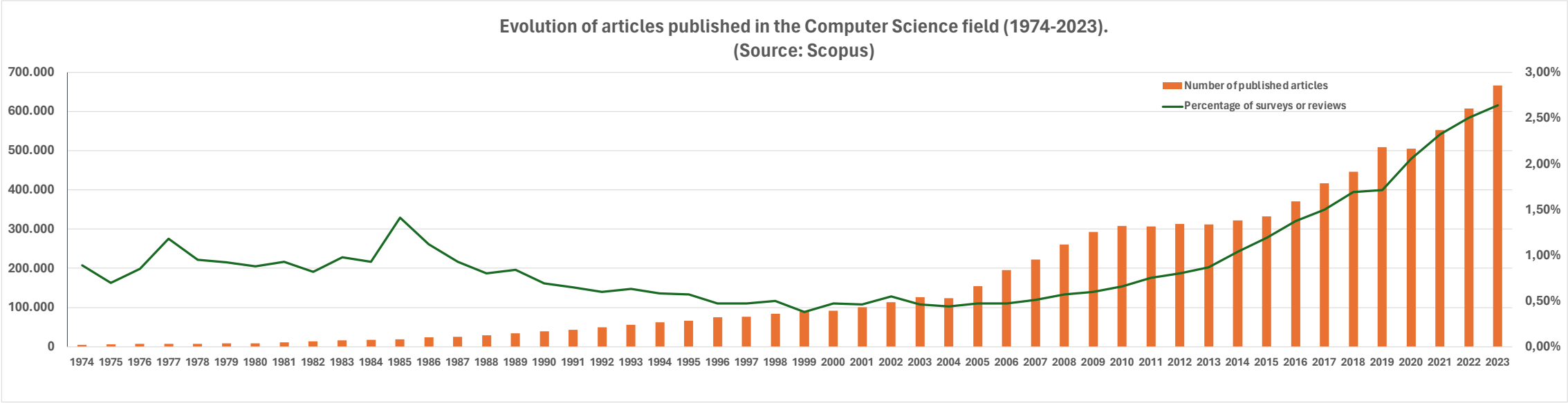

This example reflects the increasing trend of publishing surveys and reviews, driven by their advantages for authors and journals. However, this trend carries several negative consequences:

- Reviews and surveys are secondary sources that synthesize and summarize primary research. Relying heavily on them can distance new research from the original findings and introduce biases based on the author's perspectives and the selection included.
- Citing reviews instead of primary sources can overshadow the original researchers' contributions. It is crucial to give credit to those who produced the foundational insights. An illustrative case is where 64,403 citations were directed to just five review articles in the Computer Science field, rather than the original research they summarized (505 articles), each losing out on about 127 citations on average.

Academic integrity requires proper recognition of the original sources of knowledge. Overreliance on reviews or surveys can unfairly diminish the contributions of original researchers, impacting their visibility, influence, h-index, and opportunities for career advancement.

When everyone benefits from a particular practice, there may be little room for criticism. However, the widespread publication of surveys and reviews significantly distorts the scientific landscape. While it benefits certain journals and the authors of these publications, it does a disservice to the true contributors to scientific progress, whose original research is often overshadowed.

A practical solution to address the distortion caused by excessive citations of surveys and reviews is to strongly encourage authors and reviewers to prioritize citing primary research in journal articles. This would involve a shift in citation practices, where references are made directly to original studies rather than to summaries provided by reviews. Such a shift can have several benefits. For instance, it would ensure that the recognition for scientific contributions is appropriately allocated to those who conducted the foundational research, rather than those who summarized it. This practice upholds the integrity of citation metrics like the h-index, which are intended to reflect the impact of individual contributions rather than synthesized overviews.

While reviews and surveys published in journals serve a useful role in providing comprehensive overviews and helping to contextualize complex fields, their citation should be more suitable in venues like conference papers. Conference papers often face stricter length constraints, making it challenging to include extensive background information. In such cases, citing a review article can be an efficient way to acknowledge a body of literature without delving into individual studies. This selective use of reviews ensures that they fulfill their role as summary tools without overshadowing the contributions of original research.

In contrast, journal articles should focus on building upon primary research. Citing original studies in journal articles allows for a deeper engagement with the specific methodologies, findings, and interpretations that form the basis of scientific progress. This approach not only elevates the rigor of academic discourse but also prevents the inflation of citation counts for review articles that can skew the perceived quality of journals and researchers. By adhering to this practice, the scientific community can maintain the utility of reviews for educational and synthesizing purposes while ensuring that metrics like impact factor more accurately reflect substantive contributions to science.

In conclusion, adopting more discerning citation practices could foster a fairer distribution of recognition among researchers. Prioritizing the citation of primary studies in journal articles would prevent the undue elevation of review articles and restore focus on original contributions. This change would not only reinforce the validity of citation-based metrics but also encourage deeper engagement with the fundamental research that drives scientific fields forward. Upholding these principles is essential for preserving the integrity of academic evaluation systems and promoting true scientific advancement.

**References**

Ioannidis JPA. How to Make More Published Research True. PLoS Med 11(10): e1001747 (2014). https://doi.org/10.1371/journal.pmed.1001747

Garfield, E. Citation Indexes for Science. Science122,108-111(1955). DOI:10.1126/science.122.3159.108

Bao, H., Teplitskiy, M. A simulation-based analysis of the impact of rhetorical citations in science. Nat Commun 15, 431 (2024). https://doi.org/10.1038/s41467-023-44249-0